\pdfoutput=1
\documentclass[10pt,conference]{IEEEtran}
\usepackage[english]{babel}
\usepackage{titletoc}

\usepackage{amsmath}
\usepackage{graphicx}
\usepackage[colorlinks=true, allcolors=blue]{hyperref}
\usepackage{amssymb}
\usepackage[most]{tcolorbox}
\usepackage{subcaption}

\newcommand{\alert}[1]{\textbf{\color{green}
[#1]}\marginpar{\textbf{\color{green}**}}\typeout{ALERT:
\the\inputlineno: #1}}

\title{Automated Validation of LLM-based Evaluators for Software Engineering Artifacts}

\author{%
  Ora Fandina, Eitan Farchi, Shmulik Froimovich, \\
  Rami Katan, Alice Podolsky, Orna Raz, Avi Ziv \\ \\
  IBM Research, Haifa, Israel \\
  \texttt{\{ora.nova.fandina, shmulik.froimovich, alice.podolsky\}@ibm.com} \\
  \texttt{\{farchi, rami.katan, ornar, aziv\}@il.ibm.com}
}

\begin{document}

\maketitle
\begin{abstract}

Automation in software engineering increasingly relies on large language models (LLMs) to generate, review, and assess code artifacts. However, establishing LLMs as reliable evaluators remains an open challenge: human evaluations are costly, subjective and non-scalable, while existing automated methods fail to discern fine-grained variations in artifact quality.

We introduce REFINE—Ranking Evaluators for FIne-grained Nuanced Evaluation, an automated framework for benchmarking LLM-based evaluators across software engineering tasks. REFINE comprises of two modules: Hierarchy Dataset Builder applies novel generation techniques to automatically synthesize artifacts with progressively reduced quality, and Evaluator Tester quantifies each candidate evaluator configuration by measuring how closely its rankings align with expected ordering.

A key feature of REFINE is controllability: users can tune the granularity of degradation to progressively refine evaluator configurations, from coarse filtering to stress-testing on subtle quality gaps.

While the methodology is general, we focus on coding tasks reflecting the practical demands in our production setting. REFINE was integrated into IBM’s internal development workflows and applied to code generation, translation, and summarization for COBOL, an enterprise-critical programming language, using industrial data. It was used to identify LLM-as-a-Judge configurations that lifted alignment scores from below 0.7 to above 0.9 in some coding tasks. These nuance‑sensitive evaluators are now actively used by model‑training teams to support model-release decisions.

\end{abstract}

\section{Introduction}
Large Language Models (LLMs) are rapidly transforming software engineering (SE) by enabling advanced automation across tasks such as code generation, bug detection, test generation and refactoring. In addition to generating artifacts, LLMs are increasingly used as evaluators—ranking alternative implementations, identifying potential issues pull requests, and assessing generated code in end-to-end workflows. As these roles become embedded in real-world SE pipelines, ensuring the reliability and consistency of LLM-based evaluators becomes a critical challenge for industrial adoption.

LLM-based evaluators, often referred to as LLM-as-a-judge, are widely used across  domains. However, numerous studies have identified their limitations: from bias and verbosity to inconsistency and prompt sensitivity \cite{chen-etal-2024-humans}, \cite{wang-etal-2024-large-language-models-fair}, \cite{chatbot-arena}. These issues cast a doubt on their trustworthiness in business critical evaluations and pose risks when such models are used in production pipelines without rigorous validation.

In the software engineering domain, current approaches to validating LLM-based evaluators often rely  on human judgment, which is inherently subjective and expensive. Manual evaluation cannot keep pace with the rapid scale and speed at which LLM-powered tools are being integrated into modern software development environments. This highlights the urgent need for automated and scalable validation methods tailored specifically to software engineering contexts \cite{cecchini-etal-2024-holistic}, \cite{KA23, bowman-dahl-2021-will}.

Recent research has introduced automated artifact generation as a sanity check for LLM evaluators. For example, frameworks such as FBI \cite{doddapaneni-etal-2024-finding} and Drowzee \cite{LLYLWKH24} inject errors or inconsistencies into textual or code artifacts to test whether evaluators can appropriately flag them. Others, such as METAL \cite{metal-24} and MORTAR \cite{guo2025mortarmultiturnmetamorphictesting}, employ metamorphic testing to generate robust perturbations across conversational and programming contexts. 

While these efforts explore important directions in automated validation, they primarily target obvious flaws (e.g., syntax errors or factual inaccuracies) and offer limited support for evaluating nuanced quality aspects that are critical for trustworthy deployment in SE pipelines, such as code maintainability, efficiency, or adherence to design patterns. Moreover, existing methods are generally designed for mainstream languages and are not tailored to enterprise-critical legacy programming languages (such as COBOL, REXX, etc.), which require domain-specific handling.

\begin{figure*}[h!]    \centering
    \includegraphics[width=\textwidth]{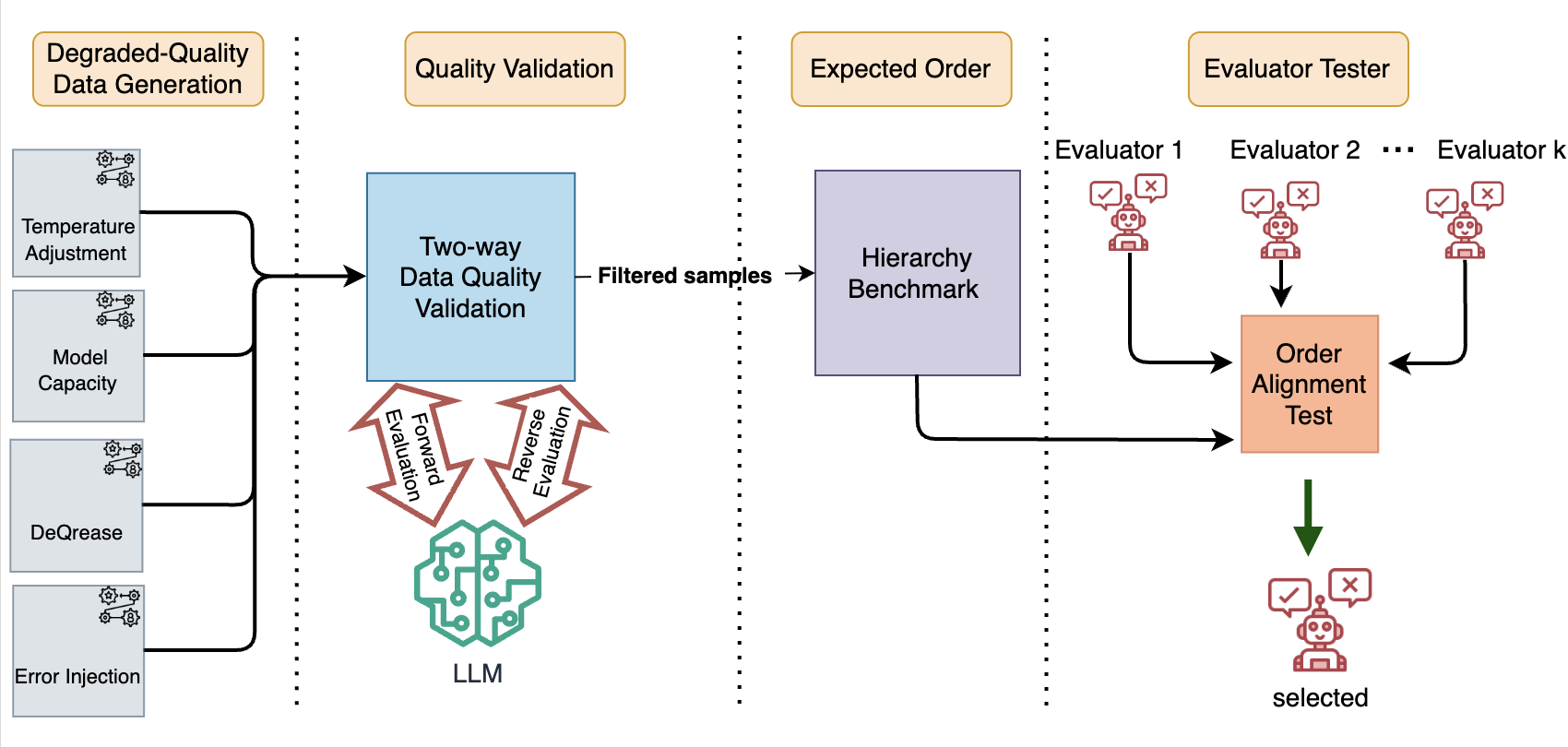}
   \caption{REFINE framework: one step in a refinement cycle. In the first stage, the most suitable generation method for a given code task is applied to construct a fine-grained hierarchy of outputs with progressively reduced quality. This hierarchy is then passed through a two-way LLM-based validation mechanism to filter out samples that do not conform to the intended quality ordering. The resulting benchmark is used in the order test to select the evaluator (judge) that best aligns with the expected ranking.}

    \label{fig:your_label}
\end{figure*}

In this paper, we present REFINE, a framework designed to rank LLM-based evaluators through fine-grained, nuanced assessment of software artifacts in an automated, scalable manner. Our approach systematically generates code artifacts with intentionally varied and controlled quality levels, producing benchmarks paired with explicit generated expectations. These benchmarks allow evaluators to be assessed for their ability to consistently rank artifacts according to expected quality hierarchies without requiring human intervention.

A key feature of our framework is fine-grained {\bf control} over the granularity of quality differences in the generated artifacts. Users can configure generation parameters to produce either coarse degradations, creating large quality gaps, or subtle, nuanced degradations that challenge more sensitive evaluators. This supports a progressive refinement strategy: initial rounds can use coarse distinctions to eliminate weak configurations, followed by increasingly refined benchmarks to identify evaluators capable of detecting subtle quality differences. This capability has proven essential in practical pipelines, where evaluator configuration is often iterative and use-case dependent.

The core idea of this paper is \textbf{automated expectation generation}: for each benchmark instance, the Hierarchy Dataset Builder produces a set of \(k\) artifact variants with progressively reduced quality. 

We developed both novel and straightforward techniques for this quality degradation process. The straightforward methods include leveraging known model strength differences and increasing decoding temperature. In addition, we introduce two novel techniques: \textit{DeQrease}, a custom decoder designed for generating lower-quality outputs in a controlled manner, and \textit{Domain-Aware Error Injection}, which introduces targeted perturbations informed by task-specific domain knowledge

While the framework supports arbitrary values of \(k\), we focus on \(k = 3\) for clarity and consistency. The resulting variants, denoted \(O_1, O_2, O_3\), are constructed to reflect a descending quality hierarchy:
\[
s(O_1) > s(O_2) > s(O_3)
\]
where $s(\cdot)$ denotes the latent quality level assigned to each variant during benchmark construction—reflecting the expected ordering a reliable evaluator should follow. To ensure the quality hierarchy holds in practice, the variants are passed through a two-way high-resolution LLM-based judgment mechanism, which filters out any samples where the intended order is not reliably recognized.  The Evaluator Tester then assesses each candidate evaluator by computing the average pairwise ordering agreement score, which reflects how consistently the evaluator preserves the expected hierarchy. Based on these scores, the best-aligned evaluators are selected to proceed to the next refinement cycle, where a new benchmark with more subtle quality differences is generated. This iterative process continues, gradually increasing granularity and refining candidates for reliable evaluator selection.

In this paper, we demonstrate how REFINE can be applied to three tasks: code translation, code summarization, and natural language–to–code generation, all focused on COBOL, an enterprise-critical language at the core of ongoing modernization efforts at IBM. We used real-world datasets derived from production COBOL systems at IBM for these tasks. 

For each task, we test 12 distinct LLM-based candidate evaluator configurations, spanning three families of LLM models (LLaMa, Mistral, and DeepSeek), and demonstrate how REFINE can be used to identify the most aligned configuration within a single iteration phase. While most tasks in this paper are illustrated using a single refinement cycle, for the code explanation task we present two full REFINE cycles. This allows us to showcase how prompt design can be iteratively improved.

Although the complete refinement process involves additional internal cycles and was used to select evaluator configurations now deployed in production, we cannot disclose the final production configurations. The use cases shown nonetheless reflect REFINE’s practical role in guiding evaluator selection under realistic industrial conditions.

COBOL presents unique challenges for LLM-based evaluators due to its specialized vocabulary, limited representation in pretraining corpora, and rigid structural conventions such as long-range control flow and declarative data definitions. These factors make it difficult for general-purpose LLMs to reason effectively over COBOL artifacts. To our knowledge, this is the first systematic study to construct fine-grained hierarchies and to benchmark LLM-based evaluators on COBOL-centered software engineering tasks grounded in real production data.

REFINE was integrated directly into IBM’s internal workflows to support evaluator selection across these tasks. 
Weak evaluator configurations were eliminated in early testing rounds, while stronger candidates underwent successive refinement cycles using progressively nuanced benchmarks. The top-performing evaluator was ultimately handed off to training and product teams to support decision-making in release readiness and quality assurance processes. \\

\noindent
{\bf Main Contributions.} The main contributions are:

\begin{enumerate}
\item  An automatic order-based testing framework for evaluating LLM-as-a-Judge (LaaJ) configurations.

\item  We demonstrate REFINE’s ability to identify high-quality evaluator configurations in a single refinement phase, for three coding tasks with real-world data.

\item  A novel decoding algorithm, \textit{DeQrease}, for fine-grained, controllable degradation that might be of interest beyond this work due to its general statistical properties.

\item   \textit{Domain-Aware Error Injection} technique for generating quality-controlled variants by introducing targeted perturbations informed by task-specific domain knowledge.

\end{enumerate}

\section{REFINE - Technical Details}
Our framework consists of two main parts: generation and validation of hierarchical data with expected degraded levels of quality, and using it for testing the order alignment of a candidate LLM evaluator with the expected result.

\subsection{Degraded‑Quality Data Generation}

\tcbset{
  imagebox/.style={
    colback=gray!10,
    colback=white,   
    boxrule=0.3mm,
    arc=2mm,
    shadow={0mm}{-1mm}{0.3mm}{black!30},
    boxsep=0mm,
    left=1mm,
    right=1mm,
    top=1mm,
    bottom=1mm,
    valign=center,       
    height=6cm,          
    width=\linewidth,    
  }
}

\begin{figure*}[t]
    \centering
    \begin{subfigure}[t]{0.3\textwidth}
        \centering
        \begin{tcolorbox}[imagebox]
        \includegraphics[width=\linewidth]{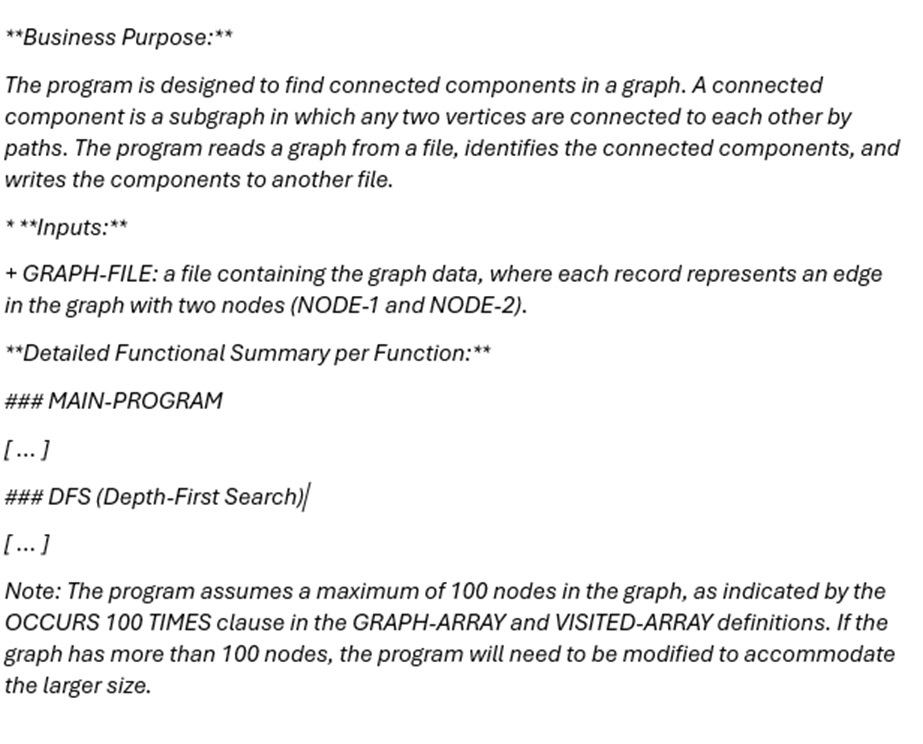}
        \end{tcolorbox}
        \caption{Summary 1 – O1 quality. The summary is fully correct, adheres to the instructions, and reflects high-quality output.}

        \label{fig:translation}
    \end{subfigure}
    \hfill
    \begin{subfigure}[t]{0.3\textwidth}
        \centering
        \begin{tcolorbox}[imagebox]
        \includegraphics[width=\linewidth]{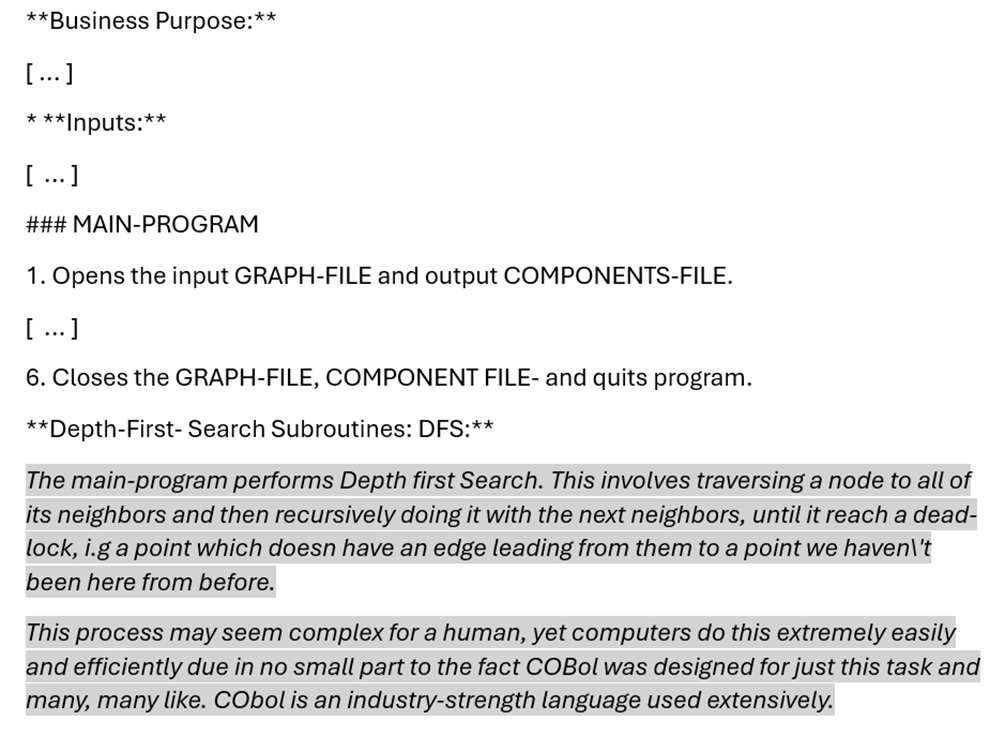}
        \end{tcolorbox}
        \caption{Summary 2 – O2 quality. Begins similarly to the first summary but degrades into general, off-task observations about COBOL. A medium level summary.}

        \label{fig:summarization}
    \end{subfigure}
    \hfill
        \begin{subfigure}[t]{0.3\textwidth}
        \centering
        \begin{tcolorbox}[imagebox]
        \includegraphics[width=\linewidth]{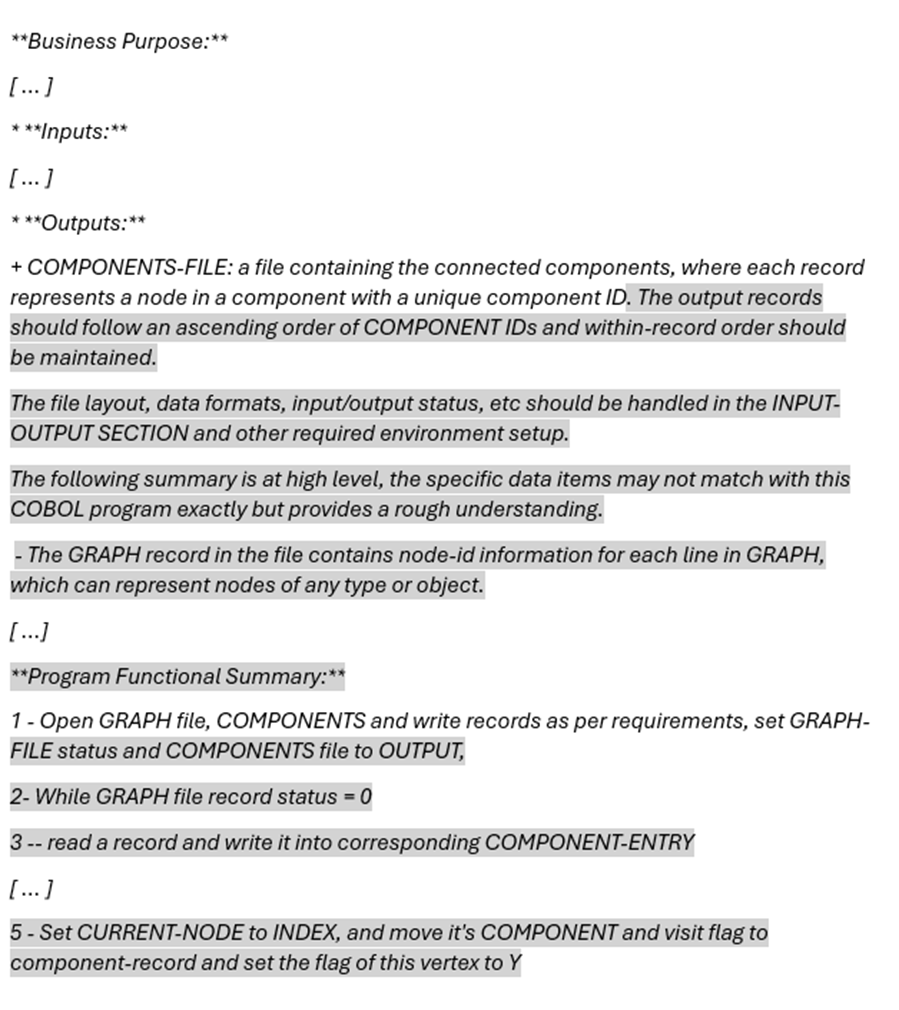}
        \end{tcolorbox}

   \caption{Summary 3 – O3 quality. Degrades quickly into an incorrect explanation, including inaccurate variable references and misrepresented functionality. A low-quality summary.}

        \label{fig:expl3}
    \end{subfigure}
\caption{Three quality levels generated by the DeQrease method for the COBOL explanation task. The input COBOL program (not shown here due to space constraints) is a synthetic example that computes connected components in a graph. For brevity, most correct details in the explanations have been omitted.}

    \label{fig:deqrease}
\end{figure*}

We introduce three automatic degradation schemes, each suited to a different class of code-related tasks. These methods use a generator LLM to produce outputs at multiple quality levels from task-specific inputs. Crucially, each scheme supports controllable granularity, enabling users to generate either large quality gaps for coarse filtering or subtle degradations for refining nuanced evaluator behavior.

\paragraph{\bf Generation with Reduced Model Capacity} Quality is degraded by generating artifacts with lower‑capacity LLMs, whose weaker representations naturally yield a lower‑quality tier relative to outputs from stronger models. In code‑translation tasks this capacity gap is particularly pronounced: large models capture cross‑language syntax and idiomatic API migration, whereas smaller models frequently emit syntactically invalid code, retain source‑language APIs, or omit corner‑case logic. 

\paragraph{\bf Decoder for Degraded Quality Generation: DeQrease}
This method uses a decoding algorithm we designed specifically for controlled quality reduction. At every generation step, the decoder restricts candidates to the top‑k tokens, then redistributes probability mass toward the lower‑ranked tokens in that set. High‑probability tokens remain selectable but are less favored, yielding outputs intentionally degraded relative to the model’s default distribution. DeQrease is especially suited to code‑summarization and explanation tasks, where “low‑quality” variants must stay coherent and relevant but may include subtle factual omissions or minor inaccuracies.

To finely control degradation severity, DeQrease introduces several tunable hyperparameters:

\begin{itemize}
\item prefix length ($0 < p \leq 1$): a fraction of the baseline (potentially high-quality) output length that is deterministically copied before degradation begins. A shorter prefix results in greater degradation, as fewer tokens are inherited from the original high-quality summary.
\item top k: sets the candidate tokens at each decoding step. Larger values extend variety, beyond the correct tokens contributing to quality degradation.
\item temperature ($t > 1$): sharpens the redistribution of probability mass, amplifying randomness and pushing generation further from the model's most likely predictions.
\end{itemize}

Together, these parameters allow DeQrease to produce controlled variants with varying fidelity while preserving relevance and syntactic fluency. An example output of DeQrease is shown in Figure~\ref{fig:deqrease}.

\paragraph{\bf Domain-Aware Error Injection}
Starting from high‑quality ground‑truth artifacts, this method uses domain knowledge to insert realistic defects without rendering the artifact nonsensical. Perturbations include changed constants, misplaced control logic, off‑by‑one errors, API misuse. The resulting variants mirror common software‑engineering issues and are especially suited to tasks such as code generation,

In addition, we designed an automated, LLM-based two-way validation scheme that evaluates each variant from both directions, assessing the quality of the output given the input, and vice versa. This phase retains only samples where both perspectives confirm a consistent decrease in quality.

\subsection{Evaluator Tester: Alignment with Expected Order}

Given a code-related task, let $\mathcal{X}$ denote its input dataset. For each input $x \in \mathcal{X}$ and an associated output $o$, an evaluator $E$ assigns a real-valued quality score $s_E(x, o)$. We place no constraints on the form or scale of these scores.

We define the ordering  alignment metric. Let $(\mathcal{X}, O_1, O_2, \ldots, O_k)$ denote the generated hierarchy of artifacts for a given task, where $\mathcal{X}$ is the set of task inputs, and $O_1, \ldots, O_k$ are the corresponding sets of degraded outputs, ordered such that the quality of outputs satisfies $O_1 > O_2 > \cdots > O_k$. 

Assume that $|\mathcal{X}| = n$ and $|O_j| = n$ for all $j$, so that each input has exactly one associated output at each quality tier.
For every input $x \in \mathcal{X}$ and its corresponding outputs $o_1, o_2, \ldots, o_k$, where $o_j \in O_j$, we evaluate the ability of an evaluator $E$ to assign scores that respect the expected quality ordering. We define:

\[
\alpha_E(x) = \frac{1}{\binom{k}{2}} \sum_{1 \leq u < v \leq k} \mathbb{I}\left[ s_E(x, o_u) > s_E(x, o_v) \right],
\]

where $\mathbb{I}[\cdot]$ is the indicator function that returns 1 when the evaluator assigns a higher score to the higher-quality output, and 0 otherwise.

The Alignment Score is obtained by averaging over all inputs:

\[
\text{Alignment Score}(E) = \frac{1}{n} \sum_{x \in \mathcal{X}} \alpha_E(x).
\]

Alignment score close to 1 indicates strong agreement with the benchmark hierarchy, suggesting that the evaluator reliably detects quality differences. In contrast, values near 0 reflect poor alignment with the expected quality ordering.

It is important to note that we intentionally adopt strict inequality in the comparison, i.e., $s_E(x, o_u) > s_E(x, o_v)$, to avoid overestimating the performance of weak evaluators that produce tied scores. For example, an evaluator that always assigns the same score (or only two distinct scores) to all outputs—regardless of their true quality—would trivially satisfy non-strict monotonicity and achieve a perfect score under a relaxed definition. By enforcing strict comparisons, we ensure that the alignment score  metric captures the evaluator's true ability to discriminate between outputs of different quality levels.


\section{REFINE in Practice: Experimental Snapshot}
We demonstrate a representative application of REFINE on three code tasks directly aligned with IBM’s production needs: code translation, code summarization, and natural language–to–code generation. While REFINE has been deployed internally in a full iterative workflow to support evaluator selection for model-training teams, the results presented in this section reflect a single evaluation phase conducted on real-world task specific COBOL data. We now describe the exact experimental setup used in this snapshot.\\

\noindent
{\bf COBOL snippets input data.} For the code translation and code generation tasks, the input datasets consist of COBOL snippets sourced from proprietary enterprise training data. 

For the code summarization task, the input COBOL snippets were synthetically generated. Unlike translation or generation, summarization does not require fully correct or executable source code. Rather, it relies on the ability to infer and express the semantic intent conveyed by the code’s structure and identifiers. As such, synthetic snippets, while potentially incomplete or imperfect, still provide a valid and diverse basis for evaluating explanation quality. This approach also enables targeted variation and inclusion of underrepresented patterns to stress-test summarization evaluators. We used the generation process introduced in \cite{farchi2024auto}.

For all tasks, we generated a three-tiered hierarchy of outputs: $O_1$ represents high-quality initial outputs, $O_2$ corresponds to medium-quality variants, and $O_3$ contains low-quality outputs. We further validated and filtered samples \((x, o_1, o_2, o_3)\) to ensure that the quality levels reflect the intended degradation. 

We applied a task-specific, automatic LLM-based two-way validation procedure to discard samples with inconsistent quality ordering. \\

\noindent
{\bf Two-way validation and filtering.} To ensure high-quality benchmark construction across tasks, we employed a two-way validation scheme using high-resolution LLM-based evaluators. These evaluators were designed to detect subtle quality differences with significantly greater sensitivity, approximately tenfold compared to the candidate LLM-as-a-Judge (LaaJ) under evaluation. This setup enables reliable quality assessment independent of the specific characteristics of the candidate LaaJ. 

Each output artifact $o_i$, $i \in \{1,2,3\}$ was evaluated in both forward and reverse directions: the forward evaluator scored the quality of the output $o_i$ given the input $x$, while the reverse evaluator assessed how well the input $x$ could be reconstructed from the output $o_i$. Both evaluators produced quality scores on a 0–100 scale. The final score for each artifact was computed as the arithmetic mean of the two directional scores. Task-specific prompts used in these evaluations and the base LLM models are provided in the relevant sections.

To enforce ordinal consistency, the averaged scores were used in a filtering step. For each input \( x \in \mathcal{X} \) and its associated set of artifacts \( o_1, o_2, o_3 \) we
verified that the average scores respected the expected hierarchy:
\[
\text{avrg\_score}(o_1) \geq \text{avrg\_score}(o_2) \geq \text{avrg\_score}(o_3).
\]
Only triplets satisfying this monotonicity criterion were retained in the final hierarchy benchmark.\\

\noindent
\textbf{Candidate LaaJs.} The resulting benchmark data was used to evaluate candidate LaaJs tailored to each task. Each evaluator assigns a scalar score from 1 to 7, with 1 denoting low-quality outputs and 7 representing high-quality outputs.
The underlying LLMs in our LaaJs use greedy decoding, as it aims to satisfy the core requirements of evaluation: determinism, stability, and consistency. 

For the evaluator LLM, we selected six models spanning three major families, used consistently across all tasks in this work:

\begin{itemize}
\item \texttt{llama-3-2-3b-instruct}, \texttt{llama-3-405b-instruct}, \texttt{llama-4-maverick-17b-128e-instruct-fp8}
\item \texttt{mistral-medium}
\item \texttt{deepseek-v3}, \texttt{deepseek-coder-33b-instruct}
\end{itemize}

Each model was evaluated under two prompt configurations, which are detailed in Appendix~\ref{app:candidate_prompts}.

The post-processing LLM used in all tasks in this work was \texttt{llama-3-70b-instruct}.

All evaluation experiments were conducted using the IBM Watsonx.ai platform.

\section{COBOL Summarization Task}
In this task, a model is given COBOL code snippets and is asked to generate a summary explaining the program. The desired summaries may vary in level of detail and focus, depending on the use case.

In our setting, the target summaries are intended to convey the business purpose of the program, its inputs and outputs, and an overall functional description.\\

\noindent
{\bf Input COBOL snippets dataset} We automatically generated $285$ synthetic COBOL code snippets, spanning various topics, difficulties, and lengths: from classic graph algorithms to small business applications \cite{farchi2024auto}.\\

\noindent
\textbf{Degraded summaries generation} 
In this task, we present a two-phase refinement scenario for evaluator selection, in which we generated two sets of degraded summaries using our DeQrease decoding method. DeQrease is specifically tailored to produce controlled-quality variants of COBOL program summaries.

In both phases, we began by generating baseline quality summaries, denoted $O_1$, using \texttt{llama-3-405b-instruct} model with greedy decoding and the following prompt:

\begin{ttfamily}
As a COBOL Expert, please provide a detailed summary of the following COBOL program, with the following sections:

1. Business purpose.

2. Inputs and outputs of the program.

3. Detailed functional summary per function.

COBOL program: \{\texttt{cobol\_code}\}
\end{ttfamily}\\

\noindent
We then produced lower-quality variants using the DeQrease method with the \texttt{granite-8b-code-instruct} model, which is trained for deep COBOL understanding. The difference between phases lies in the level of refinement in the degradation process:
\begin{itemize}
    \item In the \textbf{first phase}, we used coarser-grained degradation, with \texttt{prefix length} values set to $0.7$ for $O_2$ and $0.4$ for $O_3$, along with \texttt{top-k} = 8 and temperature $t = 7$.
    \item In the \textbf{second phase}, we generated a more refined degradation hierarchy using adjusted hyperparameters to better capture subtle quality distinctions: \texttt{prefix length}=0.8 for $O_2$, and \texttt{prefix length}=0.6 for $O_3$ quality outputs, with \texttt{top-k}=7 and temperature $t = 7$. 
\end{itemize}

\noindent
An example of the generated quality levels of the more refined second pahse appears in Figure~\ref{fig:deqrease}.\\

\noindent
{\bf Two-way validation and filtering} For both phases we used \texttt{llama-3-405b-instruct} as the forward evaluator, and \texttt{mistral-medium} as the backward evaluator, both with greedy decoding. The exact prompts are in Section ~\ref{app:ce}.

After the filtering stage 147 triplet samples, each consisting of a COBOL snippet and its corresponding summaries, were retained and included in the hierarchy benchmark $H_1$ for the first-phase data. For the second-phase data, 138 triplet samples were retained after filtering for the hierarchy benchmark $H_2$.
After the filtering stage, 147 triplet samples—each consisting of a COBOL snippet and its corresponding summaries—were retained and included in the hierarchy benchmark for the first-phase data. For the second-phase data, 138 triplet samples were retained after filtering.

To validate that the second-phase hierarchy dataset $H_2$ is indeed more refined than the first-phase dataset $H_1$, we compared their internal separation quality. We computed two quality gap metrics: 

\[
\text{Gap}_{1\to2}^{(H_1)} = \frac{\sum_{(x, o_1, o_2) \in \mathcal{X}, O_1, O_2} \text{score}(x, o_2) - \text{score}(x, o_1) }{\text {num samples in}\; H_1 }  \]

\[
\text{Gap}_{2\to3}^{(H_1)} = \frac{\sum_{(x, o_2, o_3) \in \mathcal{X}, O_2, O_3} \text{score}(x, o_3) - \text{score}(x, o_2) }{\text {num samples in}\; H_1 }  \]

The same computation was performed also for $H_2$. We obtained the following results: 

\[\text{Gap}_{1\to2}^{(H_1)}=11.82, \;\; \text{Gap}_{2\to3}^{(H_1)}=18.40
\]
and 
\[\text{Gap}_{1\to2}^{(H_2)}=7.31, \;\; \text{Gap}_{2\to3}^{(H_2)}=13.01
\]

The results confirm that $H_2$ exhibits lower quality gaps, indicating a higher level of refinement suitable for fine-grained evaluation.\\

\begin{figure}[ht]
    \centering
    \includegraphics[width=0.9\linewidth]{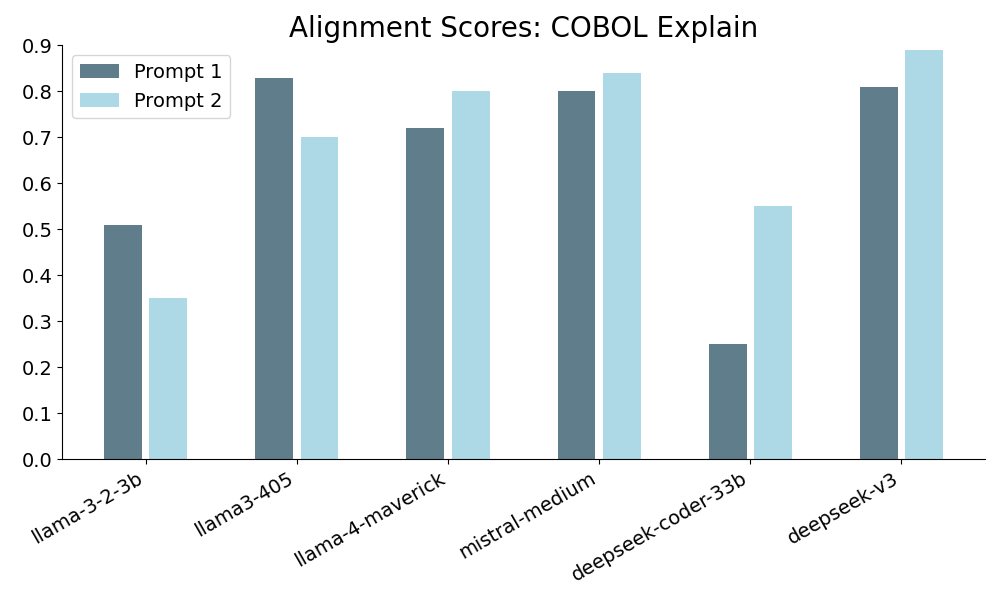}
    \caption{Alignment scores for 12 candidate evaluator configurations on the COBOL summarization task.}
    \label{fig: candidate_judges_ce}
\end{figure}

\noindent
{\bf REFINE in action: two-phase refinement,  evaluators configurations and results}
We present two phases of REFINE framework. At the first phase, we run all 12 judge configurations, as described in Candidate LaaJs paragraph, on the $H_1$ hierarchy benchmark data. The alignment scores of these first phase configurations are presented in Figure ~\ref{fig: candidate_judges_ce}.

We observe that in the explanation task the relatively weak model \texttt{llama-3-2-3b-instruct} fails the order test, while the generally strong \texttt{deepseek-coder} model performs surprisingly poorly. Additionally, although this REFINE phase allowed us to eliminate four underperforming configurations, the remaining LaaJs exhibit closely matched performance. This indicates that further refinement cycles are necessary to reliably distinguish between the top candidates.

This completes one phase of the REFINE framework. To initiate another refinement cycle, we select the top-performing LaaJ configurations from the current phase. Specifically, we chose the top three candidates to proceed: \texttt{llama-3-405b-instruct} (Prompt 1), \texttt{mistral-medium} (Prompt 2), and \texttt{deepseek-v3} (Prompt 2).

\begin{figure}[ht]
    \centering
    \includegraphics[width=0.9\linewidth]{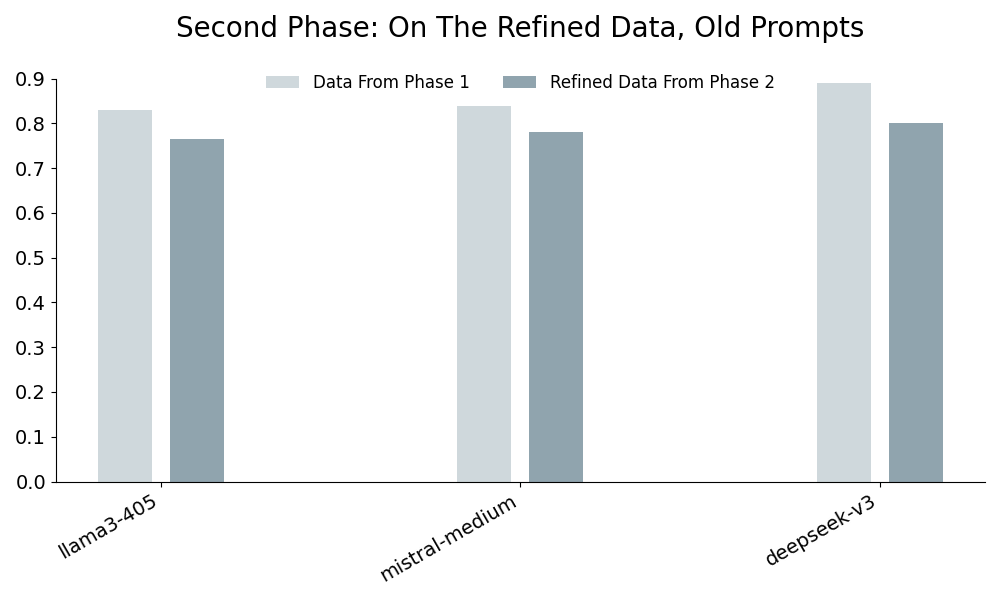}
 \caption{Alignment scores for the top three candidate configurations from Phase 1, evaluated on the original hierarchy $H_1$ and the more refined hierarchy $H_2$. The LaaJs show degraded performance on $H_2$, as expected, but remain closely matched—highlighting the need for further refinement cycles to reliably distinguish between them.}
    \label{fig:refine_data_new_old}
\end{figure}

We then evaluated these configurations on the refined $H_2$ hierarchy data and observed that, as expected, the initial LaaJ configurations struggled to capture the subtle nuances introduced in the $H_2$ data, as shown in Figure~\ref{fig:refine_data_new_old}.

\begin{figure}[ht]
    \centering
    \includegraphics[width=0.9\linewidth]{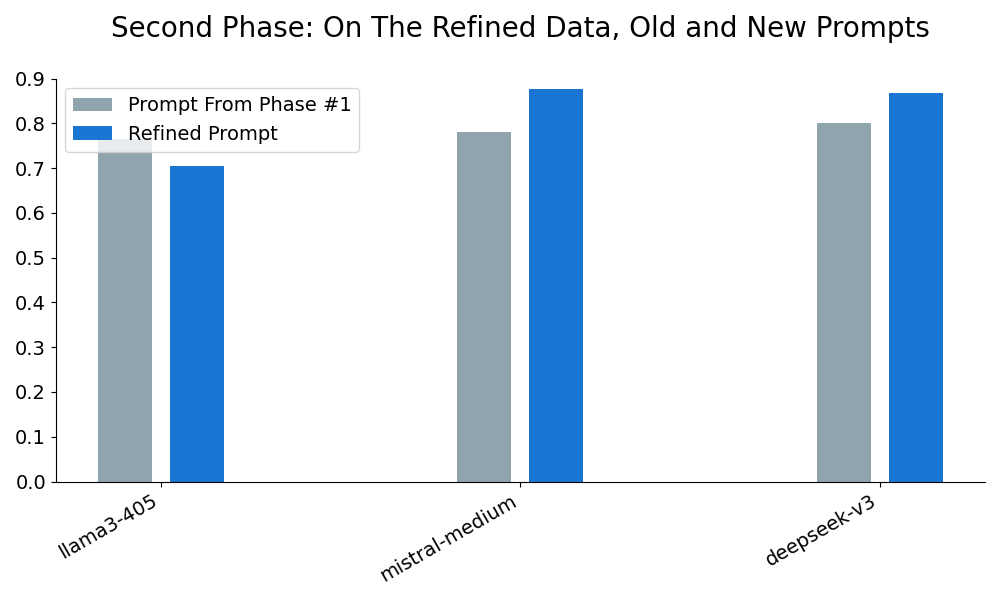}
  \caption{Top three candidate configurations from Phase 1 and their respective refined versions from Phase 2.}
    \label{fig:phase2_refine_ce}
\end{figure}

Next, we revise the prompts of the top three candidate configurations by incorporating insights from the failure cases identified in the order test. This process results in an improved prompt version, which is detailed in Appendix~\ref{app:candidate_laajs_ce}. We then evaluate the newly refined configurations on the $H_2$ dataset. The alignment scores are shown in Figure~\ref{fig:phase2_refine_ce}.

We observe that in the second REFINE cycle, two candidate LaaJs demonstrate improved alignment and can be selected to proceed to the next refinement cycle, if further evaluation is required.

\section{Code Translation: Cobol to Java}
In this task a model is instructed to translate COBOl code snippets to an equivalent code in Java language. \\

\noindent
{{\bf Input COBOL snippets dataset}}
Given the rigid structure of code translation tasks, it is crucial to use well-formed, syntactically strict COBOL code snippets, each accompanied by variable and class maps that support the translation process.

Our input dataset  $\mathcal{X}$ comprises 241 curated COBOL snippets sourced from the product’s internal evaluation suite. Originally used to assess product's performance, these samples were repurposed as foundational dataset to construct our benchmark. This ensures that LLM-evaluators capture quality variations across model outputs and remain aligned with the specific requirements of evaluating COBOL-to-code translation.\\

\noindent
{\bf Degraded translation artifacts generation}
To generate translation artifacts exhibiting varying degrees of quality degradation, we employed language models of differing capacities to convert \texttt{COBOL} code snippets into \texttt{Java}. The translation prompt is intentionally simple and directive in nature:\\

\texttt{Here is a COBOL code sample accompanied by variable and class mappings to be used in the translation.}
\texttt{Please translate the COBOL program into Java, utilizing the provided mappings.} \texttt{COBOL: \{cobol\}}

\texttt{Produce only the Java code, without any accompanying explanations or comments.}

\texttt{Enclose the resulting code between two sets of triple backticks (```)}.\\

For generating the artifact variants \(O_1\), \(O_2\), and \(O_3\), we used three models from the LLaMA-3 family, differing in parameter size, and thus in their generative capabilities:

\begin{itemize}
    \item \texttt{llama-3-70b-instruct} for higher-quality outputs 
    \item \texttt{llama-3-3b-instruct} moderately degraded outputs 
    \item \texttt{llama-3-1b-instruct} for lower-quality outputs 
\end{itemize}

A notable advantage of this approach lies in its ability to capture translation errors that arise naturally from the generative behavior of language models, as opposed to those that are artificially introduced. This allows for a more authentic and representative assessment of model performance. If an LLM-evaluator ranks the outputs in accordance with the expected quality hierarchy, it would demonstrate its capacity to distinguish between naturally occurring errors of varying severity, thereby validating its efficacy as a model evaluator.

All translations were generated deterministically using greedy decoding ,i.e. with temperature set to zero. \\

\noindent
{\bf Two-way validation and filtering}
For this phase, we used the \texttt{llama-3-405b-instruct} model with greedy decoding for both forward and backward evaluation, with the prompts that can be found in Section~\ref{app:c2j}.    

This filtering process effectively removed $67$ inconsistent samples produced during the generation step, resulting in a refined dataset of $174$ translation instances with consistently ordered quality levels.\\

\noindent
{\bf Single REFINE phase: evaluator configurations and results}

\begin{figure}[ht]
    \centering
    \includegraphics[width=0.9\linewidth]{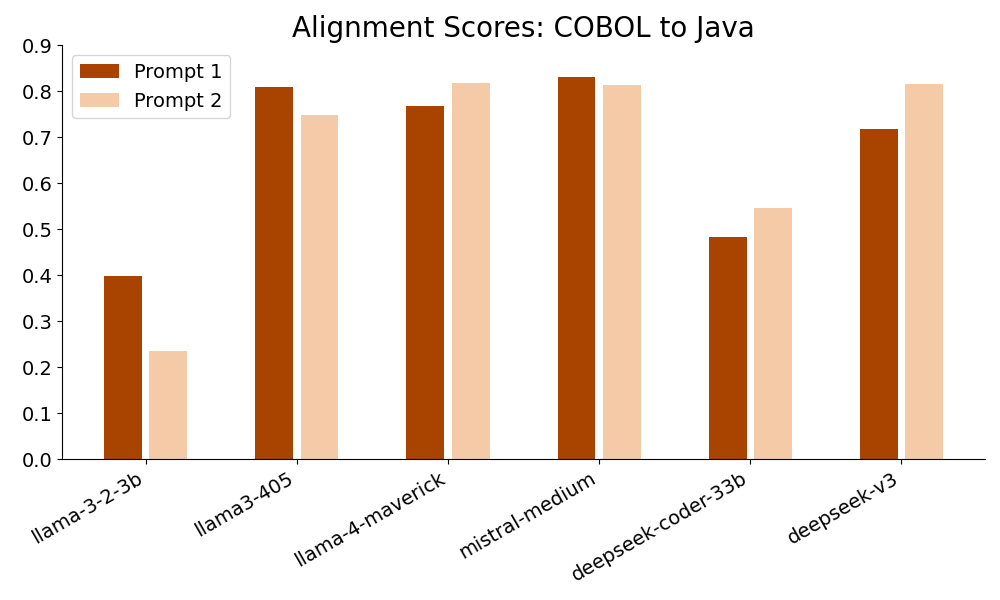}
    \caption{Alignment scores for 12 candidate evaluator configurations on the COBOL-to-Java task.}
    \label{fig:translation-judge-alignment}
\end{figure}

We applied each evaluator configuration to a degraded COBOL-to-Java dataset to measure alignment with a quality-based reference ordering. Figure~\ref{fig:translation-judge-alignment} presents the alignment scores across all candidates, illustrating significant performance variability and the utility of REFINE in identifying robust evaluators.

We observe that some configurations are closely matched and require further refinement to effectively distinguish between them and identify the most reliable LaaJ configuration. As expected, the relatively weak model \texttt{llama-3-2-3b-instruct} fails the alignment test. In contrast, the specialized and generally strong \texttt{deepseek-coder} performs unexpectedly poorly, highlighting the importance of task-specific evaluation over general model strength.

\section{Natural language instruction to COBOL}
In this task a model is required to generate COBOL code snippets from the given instruction in natural language. \\

\noindent
{\bf Input COBOLs Dataset.} For this task, we used enterprise COBOL paragraphs extracted from a real-world IBM client application that was adapted for model training process. We sampled 4208 paragraphs from this data, ranging in length from 20 to 100 lines of code.

For these high-quality, production-grade COBOL snippets, we generated natural language instructions using the \texttt{mistral-large} model, with the prompt appearing in Section ~\ref{app:nl_gen_from_cobols}. After this step, we designed two LLM-based judges for evaluating both the NL-to-COBOL task and its inverse, the COBOL-to-NL task. These judges are based on the \texttt{llama-3-70b-instruct} model and use a 1–7 evaluation scale. The evaluation prompts are 4-shot prompts; therefore, we do not disclose them in this paper. We applied these judges to all 4208 samples and retained only those for which both directional scores were 7, and samples from these 300  NL–COBOL pairs. After this filtering step, 300 COBOLs and their corresponding NL instructions were retained.  \\
 
\noindent
{\bf Degraded COBOLs Generation.}
For this task, we used the \textit{Domain-Aware Error Injection} method. We employed the \texttt{llama-3-405b-instruct} model to inject controlled errors into COBOL code, thereby lowering the quality of the (NL, COBOL) pair. To generate an additional quality level we further degraded the previously injected COBOL by introducing more errors.  
The first round of error injection used the following prompt:\\

{\ttfamily
Introduce 2 syntax or logical errors into the COBOL code.
The errors may include (but are not limited to): \\
- Introduce typos into command names (e.g., PERFROM instead of PERFORM, DISPALY instead of DISPLAY). \\
- Replace logical operators with the opposite operator (e.g., <> instead of =, < instead of >). \\
- Remove ending statements (e.g., END-IF, END-EXEC). \\
- Delete the paragraph name in a PERFORM statement.}\\

For the second level degradation we instructed the model: \\

{
\ttfamily
Locate all the names in the code (variable names, paragraph names, file names, etc.) and change half of them.

}

In this way we obtained 300 samples of an NL instruction and two associated COBOLs of regarded quality. We then pass these through two-way filtering phase. \\
 
\noindent
{\bf Two-way validation and filtering.}
For the forward pass, we used a \texttt{llama-3-405b-instruct}-based evaluator, and for the backward task, we used a \texttt{mistral-medium}-based evaluator. Both evaluators produced scores on a 1--100 scale. The exact prompts are in Appendix \ref{app:nl2c-measuring-prompts}. After this phase 234 samples retained and included in the hierarchy benchmark.\\

\begin{figure}[ht]
    \centering
    \includegraphics[width=0.9\linewidth]{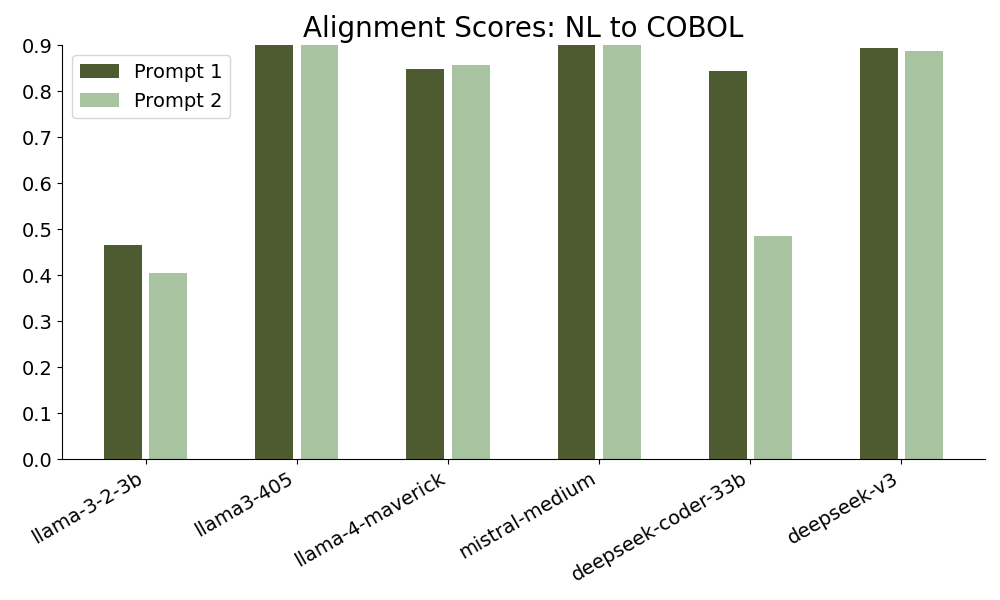}
    \caption{Alignment scores for  candidate evaluator configurations on the NL to COBOL task.}
    \label{fig:candidate_judges_nl2c}
\end{figure}

\noindent
{\bf Single REFINE Phase: Evaluator Configurations and Results}
We tested LaaJ configurations based on the same LLM models used in the previous tasks, each evaluated with two task-specific prompt variants, appearing in Appendix~\ref{app:candidate_laajs_nl2c}. The resulting alignment scores are presented in Figure~\ref{fig:candidate_judges_nl2c}.

In this task, we observe a similar phenomenon as in the previous two: the \texttt{llama-3-2-3b-instruct} model fails the order test as expected. However, the remaining models achieve high alignment scores and are closely comparable, indicating the need to proceed to the next refinement cycle to effectively distinguish between them.

\section{Conclusions}

We presented REFINE, a framework for automated validation of LLM-based evaluators in software engineering settings. REFINE combines controlled degradation with order-preserving alignment tests to support automatic and fine-grained evaluation of candidate evaluator configurations. It enables the construction of task-specific hierarchies of quality and facilitates iterative selection of the most reliable evaluators configurations.

We demonstrated the use of REFINE across three representative SE tasks: code translation, code summarization, and natural language–to–code generation, focusing on COBOL, a legacy enterprise-critical language central to ongoing modernization efforts. 

In each case, we applied REFINE to 12 candidate evaluator configurations spanning diverse model families and two prompt variants. Our experiments, grounded in production data from IBM's internal pipelines. We observed that while some weak models (e.g., \texttt{llama-3-2-3b-instruct}) are consistently discarded in early stages of REFINE, other strong models exhibit task-specific variability, underscoring the importance of domain-adapted evaluation.

The framework was integrated into IBM’s internal workflows to support evaluator selection in modernization tasks; while specific production configurations cannot be disclosed, the snapshot experiments we presented reflect realistic use cases and real data.

\bibliographystyle{IEEEtran}
\bibliography{reference}

\appendices

\section*{Prompts Book}
\addcontentsline{toc}{section}{Appendix Overview}
\startcontents[appendix]
\printcontents[appendix]{l}{1}{\setcounter{tocdepth}{2}}

\section{Filtering Phase Prompts}
All the prompts we used in two-way filtering phase are presented here.

\subsection{Cobol to Java Task} \label{app:c2j}

\noindent
{\bf Evaluation Prompt (COBOL $\rightarrow$ Java)} \\

\begin{quote}
\ttfamily
{{You are a COBOL expert.} Your task is to evaluate the Java translation of the given COBOL code snippet. Variable and class mappings are provided and should be taken into account during evaluation. Assess whether the Java code accurately reflects the original COBOL program. Provide an overall score from 0 to 100 based on the correctness of the Java translation, using the full range to reflect nuances in quality. Choose the appropriate score within each range to indicate how faithfully the translation captures the original logic.\\
\noindent
\textbf{Scoring Criteria:}
\begin{itemize}
    \item \textbf{90--100}: Fully Correct – Functionally equivalent, concise, and idiomatic translation.
    \item \textbf{75--89}: Functionally Equivalent with Minor Deviations – Accurate but verbose or non-idiomatic.
    \item \textbf{50--74}: Partially Equivalent – Mostly correct; minor issues fixable by a developer with modest effort.
    \item \textbf{25--49}: Marginally Equivalent – Some shared logic but major errors or missing features.
    \item \textbf{1--24}: Not Equivalent – Largely incorrect or functionally unrelated.
    \item \textbf{0}: Completely Incorrect – Empty, broken, unrelated, or non-functional code.
\end{itemize}

If the sample is unusable, assign a score of \texttt{-2} and terminate the evaluation. \\
\noindent
\textbf{COBOL code:} \{\texttt{given\_code}\} \\
\noindent
\textbf{Java translation:} \{\texttt{generated\_code}\}
}
\end{quote}

\noindent
{\bf \\ \\ Evaluation Prompt (Java $\rightarrow$ COBOL)}\\

{
\ttfamily
{You are a COBOL and Java expert.} Your task is to evaluate the COBOL translation of the given Java code snippet. Variable and class mappings are provided to aid in interpretation. Assess whether the COBOL code faithfully replicates the functionality of the Java code.\\

\noindent
{Scoring Criteria:}\\
Provide an overall score from 0 to 100 based on how correct the generated COBOL translation is compared to the original Java code. Use the full range of scores so that the nuances are reflected in the final score. For each score range, choose the specific score that most accurately reflects the degree of equivalence between the generated COBOL translation and the Java code.\\
Use this scale:\\
 90-100: Fully Correct-The translation is correct, functionally equivalent, concise, and idiomatic.\\
 75-89: Functionally Equivalent with Minor Deviations-The translation is accurate and functionally correct, but the COBOL code is verbose, stylistically awkward, or non-idiomatic.\\ 
 50-74: Partially Equivalent-The translation is mostly correct, but includes minor errors or omissions that can be corrected with minimal effort by a COBOL developer.\\
 25-49: Marginally Equivalent-The translation shares some logical structure with the Java code but contains significant errors, incorrect control flows, or missing features.\\
 1-24: Not Equivalent-The translation is largely incorrect, only superficially resembles the Java code, or implements unrelated functionality.\\
 0: Completely Incorrect -The translation is empty, syntactically broken, entirely unrelated, or non-executable.\\
For the selected score, provide a detailed justification. 
Include concrete examples when relevant.\\
If the sample is unusable, assign a score of \texttt{-2} and terminate the evaluation.\\
\noindent
{COBOL translation:} \{\texttt{given\_code}\} \\
\noindent
{Java code:} \{\texttt{generated\_code}\}
}\\

\subsection{COBOL Summarization Task}\label{app:ce}

\noindent
{\bf Evaluation Prompt (COBOL $\rightarrow$ Summary)}\\

\begin{quote}
\ttfamily
You are a COBOL programming expert. Your task is to carefully evaluate whether the provided summary accurately represents the given COBOL code.\\

{*** Evaluation Instructions ***}

Assign a score between 1--100, representing how well the summary matches the code. Consider the following aspects:\\

{** Special Cases:} \\
+ Assign a score of 1 if the summary is empty, completely unrelated to the code or not written in coherent English. In this case, do not proceed with the rest of the evaluation.\\

{**Content:} 
+ Deduct 5 points if the summary is not relevant to the code and its functionality. \\
+ Deduct 4 points if the summary does not accurately describe the program's purpose and main functions. \\
+ Deduct 3 points for each missing input or output file. \\
+ Deduct 3 points for each missing external dependency or interface. \\
+ Deduct 2 points for each missing detail about the program's logic or control flow. \\
+ Deduct 1 point if the summary is not concise and to the point.\\

{** Clarity:} \\
+ Deduct 5 points if the summary is not easy to understand. \\
+ Deduct 4 points if the language is not clear and concise. \\
+ Deduct 3 points for each incorrect technical term. \\
+ Deduct 2 points for each ambiguous or unclear statement. \\
+ Deduct 1 point if the summary is not well-organized and easy to follow.\\

{** Accuracy:} \\
+ Deduct 5 points for each error or inaccuracy in the summary. \\
+ Deduct 4 points for each incorrect fact or figure. \\
+ Deduct 3 points for each inconsistency or contradiction. \\
+ Deduct 2 points for each inaccurate detail. \\
+ Deduct 1 point if the summary is not up-to-date and relevant.\\

{*** Additional Considerations ***} \\
+ Deduct 2 points for each grammatical error. \\
+ Deduct 1 point for each issue with formatting or presentation. \\
+ Deduct 3 points if the summary is not comprehensive and complete.\\

{*** Scoring Guidelines ***} \\
+ Start with a score of 100. \\
+ Deduct points for each issue or error found. \\
+ Use the point values listed above as a guide.\\

{*** COBOL Code and Summary ***}

COBOL code: \{generated\_code\} \\
Summary: \{summary\}

{*** Answer ***}

Please provide the score (1--100) and a brief explanation of the issues that affected the score.
\end{quote}

\noindent
{\bf \\ \\ Evaluation Prompt (Summary $\rightarrow$ COBOL)} \\

{
\ttfamily
You are a COBOL programming expert. Your task is to carefully evaluate whether the provided COBOL code corresponds to the given summary.

*** Evaluation Instructions ***

Assign a score between 1-100, representing how accurately and completely the COBOL code matches the summary. Consider the following aspects:

** Special Cases:
+ Assign a score of 1 if the code is empty, contains only stubs, or is completely unrelated to the summary. In this case, do not proceed with the rest of the evaluation.

** Functional Coverage:
+ Deduct 5 points if the code is largely irrelevant to the purpose described in the summary.
+ Deduct 4 points if the code fails to implement the program's main goal or functionality.
+ Deduct 4 points if the summary mentions a loop or iteration over records, but code processes only a single record.
+ Deduct 3 points for each missing input or output file explicitly mentioned in the summary.
+ Deduct 3 points for each missing external dependency or interface (e.g., called programs, APIs).
+ Deduct 2 points for each supporting detail (e.g., sort step, record structure) present in the summary but not the code.
+ Deduct 2 points if the code is generic or templated and lacks domain-specific logic described in the summary.
+ Deduct 2 points for each misuse of control structures (e.g., PERFORM used when IF would be clearer, or vice versa).
+ Deduct 2 points if the code ignores a key concept in the summary (e.g., report formatting, sorting, filtering)
+ Deduct 1 point if the code is overly verbose or includes unrelated logic (i.e., lacks focus).

** Syntax and Semantics:
+ Deduct 5 points for each incorrect or illogical variable declaration (e.g., wrong PIC clauses, types).
+ Deduct 4 points for each incorrect variable usage (e.g., using a variable before initialization)..
+ Deduct 3 points for each usage of outdated or discouraged constructs like `EXIT PROGRAM` outside `MAIN`.

** Correctness and Robustness:
+ Deduct 5 points for each fatal logical bug — e.g., processing an output file as input.
+ Deduct 4 points for inconsistent naming or re-use of unrelated variable names.
+ Deduct 3 points for redundant or unreachable code blocks.

** Implementation Coverage:
+ Deduct 5 points for each hallucinated code element - something mentioned in the code that is not mentioned in the summary.
+ Deduct 4 points for each function or subprogram that is referenced but never defined or described.
+ Deduct 3 points if the summary mentions database access, but no database-related logic is implemented.
+ Deduct 2 points for each error or failure case mentioned in the summary that is not handled in the code.

*** Scoring Guidelines ***
+ Start with a score of 100.
+ Deduct points for each issue or error found.
+ Use the point values listed above as a guide.

*** COBOL Code and Summary ***

COBOL code: \{code\}, 
Summary: \{summary\}

*** Answer ***

Please provide the score (1-100) and a brief explanation of the issues that affected the score.
}

\subsection{NL to COBOL Task}\label{app:nl2c-measuring-prompts}

\noindent
{\bf Evaluation Prompt (NL $\rightarrow$ COBOL)}\\
\begin{quote}
\ttfamily
You are a COBOL programming expert. Your task is to carefully evaluate whether the provided COBOL code corresponds to the given instructions.\\

*** Evaluation Instructions ***\\

Assign a score between 1-100, representing how accurately and completely the COBOL code implements the instructions. Consider the following aspects:\\

** Special Cases:\\
+ Assign a score of 1 if the code is empty, contains only stubs, or is completely unrelated to the instructions. In this case, do not proceed with the rest of the evaluation.\\

** Functional Coverage:\\
+ Deduct 5 points if the code fails to implement the program's main goal or functionality described in the instructions.\\
+ Deduct 4 points if the instructions mention a loop or iteration over records, but code processes only a single record.\\
+ Deduct 3 points for each missing input or output explicitly mentioned in the instructions.\\
+ Deduct 3 points for each missing external dependency or interface (e.g., called programs, APIs).\\
+ Deduct 2 points for each supporting detail (e.g., sort step, record structure) present in the instructions but not the code.\\
+ Deduct 2 points if the code is generic or templated and lacks domain-specific logic described in the instructions.\\
+ Deduct 2 points for each misuse of control structures (e.g., PERFORM used when IF would be clearer, or vice versa).\\
+ Deduct 2 points if the code ignores a key concept in the instructions (e.g., report formatting, sorting, filtering).\\
+ Deduct 1 point if the code is overly verbose or includes unrelated logic (i.e., lacks focus).\\

** Syntax and Semantics:\\
+ Deduct 5 points for each usage of an "invented" statement that is not a valid COBOL statement (e.g., PERFOM, COMPUT).\\
+ Deduct 4 points for each incorrect or illogical variable declaration (e.g., wrong PIC clauses, types).\\
+ Deduct 3 points for each incorrect variable usage (e.g., using a variable before initialization).\\
+ Deduct 2 points for each usage of outdated or discouraged constructs like `EXIT PROGRAM` outside `MAIN`.\\

** Correctness and Robustness:\\
+ Deduct 5 points for each fatal logical bug — e.g., processing an output file as input.\\
+ Deduct 4 points for inconsistent naming or re-use of unrelated variable names.\\
+ Deduct 3 points for redundant or unreachable code blocks.\\

** Implementation Coverage:\\
+ Deduct 4 points if the instructions mention database access, but no database-related logic is implemented.\\
+ Deduct 3 points for each error or failure case mentioned in the instructions that is not handled in the code.\\

*** Scoring Guidelines ***\\
+ Start with a score of 100.\\
+ Deduct points for each issue or error found.\\
+ Use the point values listed above as a guide.\\

*** COBOL Code and Instructions ***\\

COBOL code: \{ \texttt{generated\_code} \} \\

Instructions: \{ \texttt{instructions} \} \\

Please provide the score (1-100) and a brief explanation of the issues that affected the score.
Do not use the character \& anywhere. Use "and" instead.
\end{quote}

\noindent
{\bf Evaluation Prompt (COBOL $\rightarrow$ NL)}\\

{
\ttfamily 
You are a COBOL programming expert.\\
You are provided with a COBOL code snippet and a set of user instructions that were generated to describe its functionality.\\
Carefully evaluate whether the user instructions accurately describe the given COBOL code.\\
*** Evaluation Instructions ***\\
Assign a score between 1-100, representing how well the summary matches the code. Consider the following aspects:\\
** Special Cases:\\
+ Assign a score of 1 if the summary is empty, completely unrelated to the code or not written in coherent English. In this case, do not proceed with the rest of the evaluation.\\
** Content:
+ Deduct 5 points if the instructions do not fully and correctly describe the COBOL code.\\
+ Deduct 4 points if the instructions do not accurately describe the program's main functions.\\
+ Deduct 3 points for each missing input or output file.\\
+ Deduct 3 points for each missing external dependency or interface.\\
+ Deduct 2 points for each missing detail about the program's logic or control flow.\\
+ Deduct 1 point if the instructions are not concise and to the point.\\
** Clarity:\\
+ Deduct 5 points if the instructions are not easy to understand.\\
+ Deduct 4 points if the language is not clear and concise.\\
+ Deduct 3 points for each incorrect technical term.\\
+ Deduct 2 points for each ambiguous or unclear statement.\\
** Accuracy:\\
+ Deduct 5 points for each error or inaccuracy in the instructions.\\
+ Deduct 4 points for each incorrect fact or figure.\\
+ Deduct 3 points for each inconsistency or contradiction.\\
+ Deduct 2 points for each inaccurate detail.\\
*** Additional Considerations ***\\
+ Deduct 3 points if the instructions are not comprehensive and complete.\\
+ Deduct 2 points for each grammatical error.\\
+ Deduct 1 point for each issue with formatting or presentation.\\
*** Scoring Guidelines ***\\
+ Start with a score of 100.\\
+ Deduct points for each issue or error found.\\
+ Use the point values listed above as a guide.\\
*** COBOL Code and Instructions ***\\
COBOL code: \{\texttt{code}\}\\
Instructions:  \{\texttt{instructions}\}\\

Please provide the score (1-100) and a brief explanation of the issues that affected the score.
}

\section{Candidate LaaJs Prompts} \label{app:candidate_prompts}
\subsection{COBOL to Java Task}\label{app:candidate_laajs_c2j}
\noindent
{\bf Prompt 1:}
{
\ttfamily
\noindent
Here is a COBOL program and Java code translation of the COBOL program.
The COBOL program includes code as well as variable and class mappings to use in the translation.\\
Please assign a score to the correctness of the Java code translation.
Use a scale from 1 to 7:\\
7 - Accurately and faithfully translates the COBOL program into Java, preserving logic and functionality; correct declarations, conditionals, idiomatic syntax, and no unnecessary changes.\\
6 - Mostly correct translation with only minor or subtle issues. \\
5 - Incomplete or logically flawed translation; basic structure and some logic preserved, but major correctness issues exist.\\
4 - Significant errors prevent compilation or correct execution; multiple logical or syntactic flaws.\\
3 - Undeclared variables, incomplete translation.\\
2 - Lacks implementation of required logic; misunderstanding of COBOL concepts.\\
1 - Empty or nearly empty method; no meaningful logic implemented.
Provide justification for the score. Begin with \#\#\#Reasoning and end with \#\#\#End\_Reasoning.\\
COBOL Program: \{{cobol\}}\\
Java Translation: \{{java}\}
}\\

\noindent
{\bf Prompt 2:}
{
\ttfamily
Here is COBOL program and a Java translation of the COBOL program. 
The COBOL program also includes variables and class mapping to use in the translation.  
Assign a score of 1-7 to the correctness of the Java translation. Where 7 is the best and 1 is the worst.
}

\subsection{COBOL summarization task}\label{app:candidate_laajs_ce}
\noindent
{\bf {Prompt 1}}:
{
\ttfamily
You are a programming expert and a very strict evaluator. Please evaluate the provided ""Source Code"" and ""Provided Summary"" by following this step-by-step process. \\
Step-by-Step Evaluation Process: \\
Understand the Source Code: Carefully read and understand the provided source code. Identify the key functionalities, structures, and logic implemented.\\
Read the Provided Summary: Read the provided summary thoroughly. Note down the main points it makes about the code's functionality and structure. If the summary uses the remarks from the Source Code that is ok.\\
List Correctness Issues:  Compare the summary with the source code.
If not perfect, List any discrepancies, inaccuracies, or missing information in the summary.
Note if the summary includes information not present in the source code than it is not considered good summary and it should be penalized heavily. \\
Assess Correctness Quality:
Evaluate how well the summary reflects the actual functionality and structure of the code.
Consider if any inaccuracies or extraneous details are present in the summary to evaluate.\\
Score Assignment:
List Completeness Issues:
If not perfect, list any missing parts in the code that the summary did not address. \\
List Conciseness Issues:
If not perfect, list any issues in the conciseness of the summary, list references to and verbose and repetitive parts in the summary.\\
Based on your assessment, give an over all score between 1 and 7. \\
Use the following scoring guidelines: \\
The score should be penalized if the ""Provided Summary"" is not directly relevant to the  ""Source Code"" . For example, A score of 1 indicates the summary is incorrect or summary is repeating code instead of text. A score of 3 indicates that summary  contains more than 1 inaccurate information which is not supported by the ""Source Code"". A score of 5 indicates that summary has most of the accurate information but contain at least 1 inaccurate information. While a score of 7 suggests it accurately describes the Source code functionality and structure.
Penalized if the ""Provided Summary"" doesn't contain all information from ""Source Code"" . A score of 1 indicates the summary doesn't capture information from Source code. A score of 3 indicates that summary has few information but also missing few information from the source code. A score of 5 indicates that summary has most of the information from Source code but missed out  at least  1 information. While a score of 7 suggests describes the Source code functionality and structure completely.
Penalize if the ""Provided Summary"" is verbose and repetitive . For example, A score of 1 indicates the summary is overly verbose or badly repetitive or may contain source. A score of 4 indicates that summary is bit verbose. A score of 5 indicates that summary is most concise.  While a score of 7 suggests that the summary is perfectly concise and cannot be reduced any further.\\
Based on your assessment, give an over all score between 1 and 7. \\
code: \{ {code} \}  
summary: \{summary\} 
}\\

\noindent
{\bf Prompt 2}:
{
\ttfamily
You are a COBOL expert.
Please evaluate the provided COBOL code and the corresponding code summary using the following criteria:\\
Correctness - assess whether the summary accurately reflects the functionality of the code.\\
Completeness - assess whether the summary provides a complete representation of the code.\\
Based on this assessment, assign an overall score between 1 and 7, where 1 reflects a very poor summary and 7 reflects an excellent.\\
COBOL code: \{code\} Summary: \{summary\}
}\\

\noindent
{\bf Refined Prompt From Phase 2}:

\noindent
{ \ttfamily

You are a COBOL programming expert. Your task is to carefully evaluate whether the provided summary accurately represents the given COBOL code.\\

*** Evaluation Instructions ***\\

Assign a score between 1.0 - 7.0, representing how well the summary matches the code. Consider the following aspects:\\

** Special Cases:\\
+ Assign a score of 1 if the summary is empty, completely unrelated to the code or not written in coherent English. In this case, do not proceed with the rest of the evaluation.\\

** Content:
+ Deduct 5\% if the summary is not relevant to the code and its functionality.\\
+ Deduct 4\% if the summary does not accurately describe the program's purpose and main functions.\\
+ Deduct 3\% for each missing input or output file.\\
+ Deduct 3\% for each missing external dependency or interface.\\
+ Deduct 2\% for each missing detail about the program's logic or control flow.\\
+ Deduct 1\% if the summary is not concise and to the point.\\

** Clarity:\\
+ Deduct 5\% if the summary is not easy to understand.\\
+ Deduct 4\% if the language is not clear and concise.\\
+ Deduct 3\% for each incorrect technical term.\\
+ Deduct 2\% for each ambiguous or unclear statement.\\
+ Deduct 1\% if the summary is not well-organized and easy to follow.\\

** Accuracy:\\
+ Deduct 5\% for each error or inaccuracy in the summary.\\
+ Deduct 4\% for each incorrect fact or figure.\\
+ Deduct 3\% for each inconsistency or contradiction.\\
+ Deduct 2\% for each inaccurate detail.\\
+ Deduct 1\% if the summary is not up-to-date and relevant.\\

*** Additional Considerations ***\\
+ Deduct 2\% for each grammatical error.\\
+ Deduct 1\% for each issue with formatting or presentation.\\
+ Deduct 3\% if the summary is not comprehensive and complete.\\

*** Scoring Guidelines ***\\
+ Start with a score of 7.0.\\
+ Deduct points for each issue or error found.\\
+ Use the point values listed above as a guide.\\

*** COBOL Code and Summary ***\\

COBOL code: {generated\_code}\\

Summary: {summary}\\

*** Answer ***\\

Please provide the score (1.0 - 7.0) and a brief explanation of the issues that affected the score.

}

\subsection{NL to COBOL task}\label{app:candidate_laajs_nl2c}

\noindent
{\bf Prompt 1}: You are a Cobol expert.
{\ttfamily
Task:
You are provided with user instructions describing COBOL code and the corresponding COBOL code generated by an AI model.
Your Goal:
Carefully evaluate whether the provided COBOL matches the user instructions.\\
Instructions for Evaluation:
Clearly state if the generated COBOL code fully matches the described requirements.\\
If discrepancies or errors exist, explicitly highlight each one, providing concise explanations and specific references to both the instruction and relevant COBOL code snippets.\\
Key points for problem searching and reporting:\\
- The code must adhere to the instructions, but no more than that. For example, if the instructions do not mention input validation or error-handling, the program may lack these features. Do not consider or comment about issues that fall outside the scope of the instructions.\\
- Sometimes, the instructions are ambiguous and can be interpreted in several ways. If the code is aligned with a common and reasonable implementation, then the code follows the instructions, even if you think that there is a more common interpretation for the user instructions. Do not consider such case as problematic.
- The code does not have to comprise a full program, do not decrease score for missing variable and procedure definitions.\\
Assign an overall accuracy score from 1 to 7 according to the next guidelines:\\
1 (Empty) - when there is no code, only text or empty output\\
2 (Irrelevant) - when the code is not related to the instructions\\
3 (Partially relevant) - when the code is mostly incorrect, but some aspects of it are positive and with an effort it can be modified to a code that adhere the instructions\\
4 (Relevant but Poor) - when the code is problematic, but mostly follows the instructions\\
5 (Fair) - when the code is mostly ok, excluding a few problems to fix to make it adhere the instructions\\
6 (Good) - when the code generally follows the instructions, where minor correctness may will make it fit the instructions perfectly\\
7 (Excellent) - when the code is entirely correct and fully meet the instructions\\
Evaluation Format:\\
Functional Correctness:\\
- Issues found (if any):\\
  1. description: ...\\
  (Repeat as needed)\\
Overall Accuracy Score: [1–7]\\
user instructions: \{instructions\}\\
COBOL code: \{code\}
}\\

\noindent
{\bf Prompt 2}:
{
\ttfamily
You are a Cobol expert.}
{\ttfamily
You are provided with user instructions describing COBOL code and the corresponding COBOL code.\\
Evaluate how well does the COBOL code implement the instructions.\\
Assign an overall accuracy score from 1 (worst) to 7 (best) to reflect much the COBOL code matches the instructions.
User instructions: \{instructions\}\\
COBOL code: \{code\}
}

\section{NL Instructions for COBOLs}\label{app:nl_gen_from_cobols}
We started from high-quality, production grade 500 COBOL paragraphs, for which we generated NL instructions by instructing \texttt{mistral-large}  LL model with the following prompt:

{
\ttfamily
You are a Cobol expert. Task:\\
You are provided with COBOL code generated by an AI model from user instructions.\\
\noindent
Your Goal:\\
Analyze what the cobol code does and create a set of instructions which a user could have used to create the cobol code.\\
The instructions should describe the code's logic, but they should not describe each single line.
If the code does something indirectly (e.g. by calling a PERFORM statement), mention that in the instructions.\\
Do not mention the exact content of print message, just their general meaning.\\
\noindent
Answer format:\\
Start your answer with "Generate Cobol code" and then a high level description of the program and a more detailed description with variable names, file names and table names afterward.\\
Do not write anything other than the instructions.
Do not write in the instruction anything that is not explicitly mentioned in the cobol code.\\
COBOL code:
\{\texttt{code}\}
}

\end{document}